\documentclass{PoS}

\title{Global fits of scalar singlet dark matter with GAMBIT}

\ShortTitle{Global fits of scalar singlet dark matter with GAMBIT}

\author{\speaker{Jonathan M. Cornell} {\rm on behalf of the GAMBIT collaboration}\\%
      McGill University\\
      E-mail: \email{cornellj@physics.mcgill.ca}}

\abstract{The wide range of probes of physics beyond the standard model leads to the need for tools that combine experimental results to make the most robust possible statements about the validity of theories and the preferred regions of their parameter space. Here we introduce a new code for such analyses: GAMBIT, the Global and Modular BSM Inference Tool. GAMBIT is a flexible and extensible framework for global fits of essentially any BSM theory. The code currently incorporates direct and indirect searches for dark matter, limits on production of new particles from the LHC and LEP, complete flavor constraints from LHCb, LHC Higgs production and decay measurements, and various electroweak precision observables. Here we present an overview of the code's capabilities, followed by preliminary results from scans of the scalar singlet dark matter model.}

\FullConference{38th International Conference on High Energy Physics\\
		  3-10 August 2016\\
		 Chicago, USA}

\begin{document}

\section{Introduction}

Theories of particle physics beyond the standard model (BSM) predict a range of experimental signatures. New particles can be produced in colliders, the rate of rare processes can be modified, and dark matter (DM) can lead to signals both in the sky and the laboratory. The results from each of these probes are often analyzed separately to determine how they constrain a particular new physics theory; however, to fully understand how constrained a theory is, there is a need for analyses in which these results are combined in a consistent way. GAMBIT, the Global and Modular BSM Inference Tool\footnote{http://gambit.hepforge.org}, is a new software package to do just this.

In a global fit, a composite likelihood function is generated based on multiple experimental results and then scanned over to determine which region of the theory parameter space the data prefer.
Tools currently exist to carry out this procedure for BSM models (\textit{e.g.} \cite{Strege15,Fittinocoverage,MasterCodeMSSM10,Butter:2015fqa}), and GAMBIT builds on these previous efforts with its focus on modularity and flexibility. GAMBIT has been designed from the beginning to be a model-independent framework for global fits, with the initial release containing the necessary tools to scan a wide range of parameterizations of the MSSM as well as a simple scalar singlet DM model.  There are a range of observables and likelihoods that will be included in the first release, and the GAMBIT framework easily allows for any of these to be enabled or disabled in a particular scan. The code has been designed to be easily extensible, so that a user can add observables, likelihoods, and models of their own. GAMBIT includes interfaces to the most commonly used computer tools in HEP to aid in its calculations, and it is easy for a user to add interfaces to other codes if they are needed.

In the following pages, we give a brief overview of the structure of GAMBIT, and then discuss in more detail how it calculates collider and dark matter constraints. We then present some preliminary results for the scalar singlet DM model that have been determined using the code.

\section{Code Structure}

\begin{figure}
\centering
\includegraphics[width=0.98\linewidth]{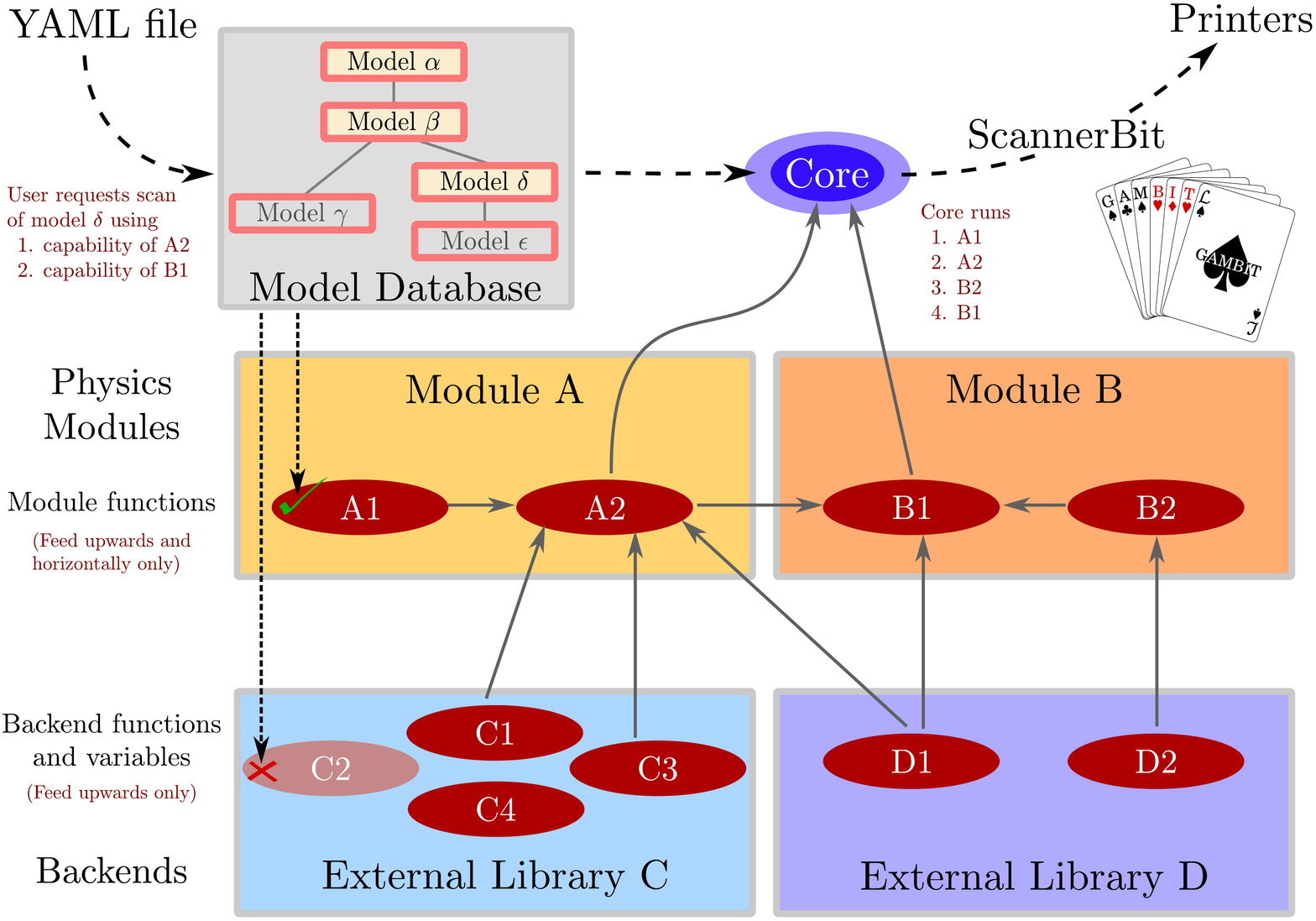}
\caption{\label{structure} A diagram showing the components of GAMBIT. The arrows show the ways the elements of the code communicate.}
\end{figure}

A schematic overview of GAMBIT is shown in Fig.~\ref{structure}. GAMBIT is composed of modules, or ``Bits", of which there are currently six physics modules and ScannerBit, which directs the scan of the model parameter space. The physics modules either calculate observables and likelihoods internally or make use of external codes (referred to as ``backends") to determine these quantities. ScannerBit chooses the parameters to be sampled using either internal algorithms or external codes 
through a plugin interface. The currently available physics modules are:
 
\begin{itemize}

\item \textbf{ColliderBit:} Calculates particle collider observables and likelihoods based on searches for new particles from the LHC and LEP.

\item \textbf{FlavBit:} Calculates likelihoods from a range of flavor physics results, primarily decay rates and angular observables for $B$ decays as observed at LHCb and $B$ factories.

\item \textbf{DarkBit:} Calculates DM observables and likelihoods, including results from direct and indirect searches and relic density constraints. 

\item \textbf{SpecBit:} Interfaces to external mass spectrum calculators to provide GAMBIT with pole masses and running parameters.

\item \textbf{DecayBit:} Calculates decay rates of BSM particles and contains SM particle decay information.

\item \textbf{PrecisionBit:} Calculates BSM corrections to a range of precision observables, including such quantities such as the mass of the $W$ and the muon anomalous magnetic moment, and provides likelihoods for these values.

\end{itemize}

Another integral component of GAMBIT is the model database. This is designed to be hierarchical, allowing for translations of constrained parameterizations of a model to more general forms (\textit{e.g.} the model database will easily allow the user to convert a 7 parameter version of the MSSM to the 63 parameter version). This allows for easy use of codes that require a certain parameterization of a model.

\section{ColliderBit}

ColliderBit provides GAMBIT with the ability to constrain BSM theories based on results of direct searches for production of new particles. This module currently contains results from a range of SUSY and dark matter (monojet) searches by both ATLAS \cite{Aad:2008zzm} and CMS \cite{Chatrchyan:2008aa} using the Run 1 dataset, and there are plans to include updated Run 2 versions of some of these searches in the first GAMBIT release. Also included in the module is the capability to recast LEP direct sparticle searches for a general SUSY model, as well as the capability to generate a likelihood for a model in light of LHC measurements of the SM Higgs parameters and limits on extended Higgs sectors, using the HiggsBounds and HiggsSignals \cite{HiggsSignals} codes as backends.

In determining the expected signal for a particular model, ColliderBit goes through the standard process of calculating a cross section, generating Monte Carlo events, and then running those events through a detector simulation. As this is a task that can be very time consuming, particularly since with GAMBIT we are interested in scanning large parameter spaces, ColliderBit is optimized to do this as quickly as possible. By default the SUSY cross sections are calculated using Pythia8 \cite{Sjostrand:2014zea}, rather than slower NLO tools, and the event generation is done using a parallelized and simplified version of Pythia8. Detector simulation is done using BuckFast, a specially written tool based on four-vector smearing.

\section{DarkBit}

In addition to the collider constraints, DarkBit adds the ability to calculate likelihoods and observables related to the direct and indirect detection of DM to GAMBIT. For indirect detection, GAMBIT includes a new code called gamLike, which allows DarkBit to calculate the likelihood of a DM model based on gamma-ray observations of a variety of targets, including dwarf spheroidal galaxies \cite{LATdwarfP8} and the galactic center \cite{Calore:2014xka} with the Fermi-LAT. DarkBit can also calculate the rate of high energy neutrinos that would be expected from DM annihilation in the sun and then determine a likelihood based on IceCube observations using an interface to the nulike package \cite{IC79_SUSY}.

For the above observables, DarkBit calculates the expected flux of particles from DM annihilation using information from a structure called the process catalog, which contains information on the rate of annihilation and decay rates of BSM particles to all possible final states. If an annihilation contains a BSM particle in the final state, DarkBit simulates the decays of that particle on the fly using the internal Fast Cascade Monte Carlo code. The information in the process catalog can also be used by DarkBit to calculate the DM relic density.

Finally, DarkBit has the ability to calculate signal rates and likelihoods for direct searches using the DDCalc code which will also be released with GAMBIT. DDCalc includes a range of direct detection results, including the most recent LUX \cite{LUXrun2} and PandaX \cite{PandaX2016} analyses. 

\section{Scalar Singlet DM}

One of the most simple models of dark matter in the literature consists of a real scalar field $S$ which transforms as a singlet under all standard model gauge groups (for more details and previous comprehensive studies of this model, see \textit{e.g.} \cite{Cline13b}). In this model interactions between the DM $S$ and standard model particles are mediated solely by the Higgs, so the phenomenology is entirely controlled by the DM mass $M_S$ and coupling to the Higgs $\lambda_{hS}$.

\begin{figure}
\centering
\includegraphics[width=0.463\linewidth]{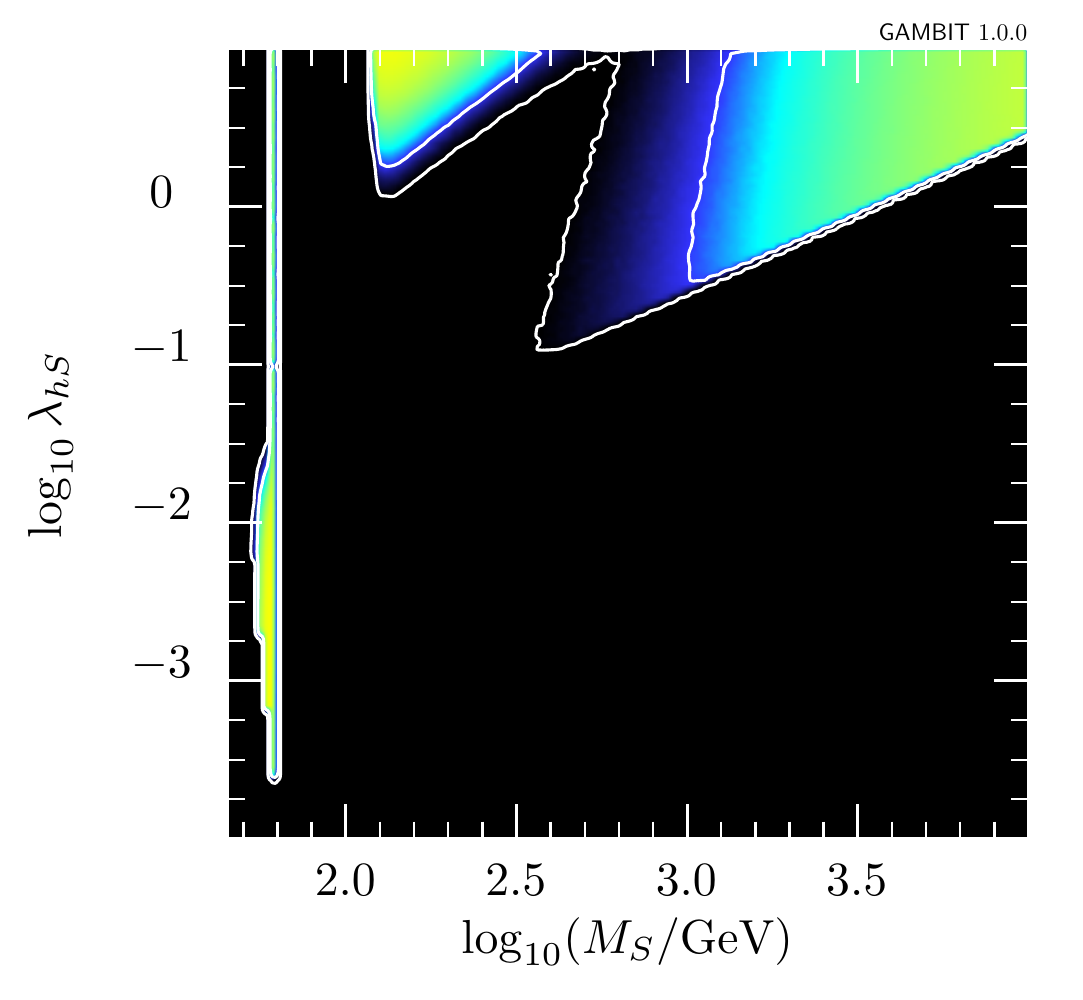}
\includegraphics[width=0.527\linewidth]{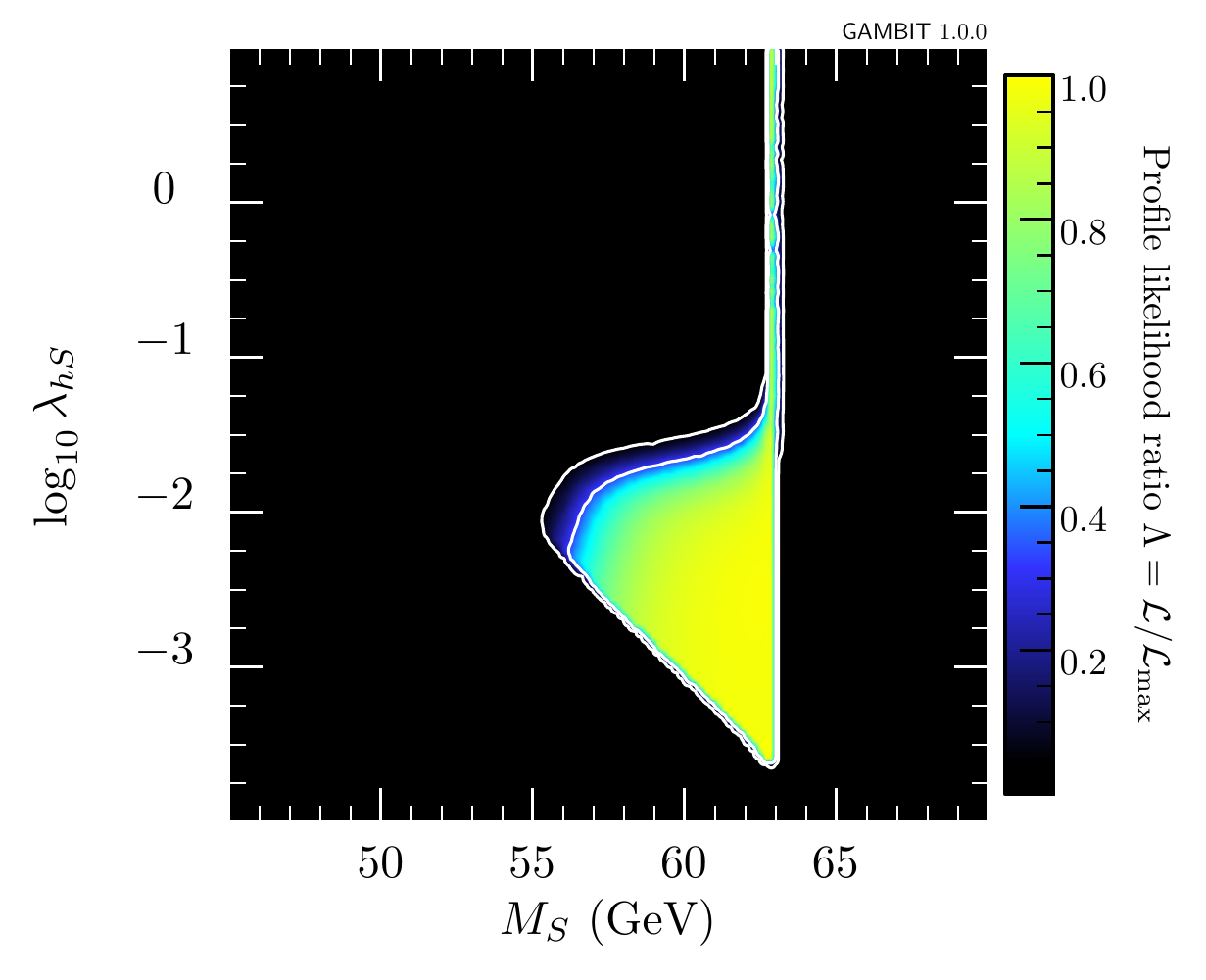}
\caption{\label{scalarDM}Plots showing the profiled composite likelihood in the parameter space of the scalar singlet DM model. The white lines surround the 68\% and 95\% credible regions.}
\end{figure}

In Fig.~\ref{scalarDM}, we present results of scans over the parameter space of this model. We have scanned the parameter space using Diver, a differential evolution scanner which will be released with ScannerBit, varying $M_S$ over the range 45 -- 10000 GeV and $\lambda_{hS}$ over the range 0.0001 -- 10. In the scan we have also varied a range of nuisance parameters that could affect the calculations, including the masses of all six quarks, the Higgs mass, $\sigma_s$ and $\sigma_l$ describing the quark content of nucleons, the local dark matter density, the Fermi coupling, and the strong and electromagnetic couplings (the strong coupling enters into loop corrections to the annihilation rate). In total, this is a 15 dimensional parameter space, and to make the plots in Fig.~\ref{scalarDM}, we have profiled over all of the nuisance parameters. The likelihood function contains contributions from the solar neutrino and dwarf spheroidal gamma ray limits, constraints from DM direct detection, limits on the invisible width of the Higgs, and the constraint that the dark matter relic density not exceed the value determined by the Planck collaboration \cite{Ade:2015xua}, as well as likelihoods for the nuisance parameters. Note that we rescale the dark matter density in the dwarfs and local Galaxy based on the calculated relic density for each model point.

The model is most strongly constrained by the need for DM to not overclose the universe. This excludes small coupling values except for when $M_S \sim 63$ GeV, when the dark matter self-annihilation rate is resonantly enhanced. Low values of $M_S$ are excluded by measurements of the Higgs invisible width, while the LUX and PandaX 2016 results exclude a large portion of the parameter space from $M_S$ = 100 GeV to $M_S$ = 1000 GeV which was previously allowed (this was also recently pointed out in \cite{He:2016mls} and \cite{Escudero:2016gzx}). In the right hand plot of Fig.~\ref{scalarDM}, we show the small coupling region which is allowed because of the resonance. This region is constrained from above by both direct detection and the Higgs invisible width. These plots show that, with the recent results from LUX and PandaX, the scalar singlet DM model is on the edge of viability.

\section{Conclusions}
In this conference paper, we have presented a brief overview of the GAMBIT code, as well as constraints on the scalar singlet DM model that have been determined using it. Scans of the more complicated MSSM parameter space are currently in progress. The results of all scans, along with a series of papers describing the GAMBIT code and its constituent modules, will soon be released, with public release of the code soon to follow.

\acknowledgments JMC gratefully acknowledges support from the NSERC.

\bibliographystyle{jhep}
\bibliography{GAMBIT-inspire}

\end{document}